\documentclass[letterpaper,prl,aps,showpacs,floatfix,twocolumn]{revtex4}
\usepackage{graphicx}
\usepackage[ansinew]{inputenc}
\usepackage{array}
\usepackage{color}
\usepackage{amsmath}
\usepackage{amsxtra}
\usepackage{amstext}
\usepackage{amssymb}
\usepackage{latexsym}
\usepackage{dsfont}
\usepackage{verbatim}

\begin{document}
\title{Quantum state transfer between a Bose-Einstein condensate \\ and an optomechanical mirror}
\author{S. Singh, H. Jing, E. M. Wright and P. Meystre}
\affiliation{B2 Institute, Department of Physics and College of Optical
Sciences\\The University of Arizona, Tucson, Arizona 85721}

\begin{abstract}
In this paper we describe a scheme for state transfer between a trapped atomic Bose condensate and an optomechanical end-mirror mediated by a cavity field. Coupling between the mirror and the cold gas arises from the fact that the cavity field can produce density oscillations in the gas which in turn acts as an internal Bragg mirror for the field.  After adiabatic elimination of the cavity field we find that the hybrid system of the gas and mirror is described by a beam splitter Hamiltonian that allows for state transfer, but only if the quantum nature of the cavity field is retained.
\end{abstract}
\maketitle

Cavity optomechanics is rapidly developing into a major area of research. Several groups have now achieved cooling of the center-of-mass motion of micromechanical systems close to the ground state \cite{OConnell2010, Teufel2011, Chan2011}, and the coherent exchange of excitation between phonons and photons characteristic of the strong coupling regime has also been demonstrated \cite{Groeblacher2009, OConnell2010, TeufelSC2011}.  In a parallel development, Bose-Einstein condensates (BECs) trapped inside high-$Q$ optical resonators have been shown to behave under appropriate conditions much like optically driven mechanical oscillators, offering an alternative, ``bottom-up'' route to study the optomechanical properties of mesoscopic systems \cite{Brennecke2008, Murch2008, Purdy2010, Brahms2010, Kanamoto2010}. Of particular interest are hybrid systems consisting of mechanical systems in the quantum regime coupled to atoms \cite{Treutlein2007, Hammerer2009, Hunger2010, Steinke2011, Camerer2011}, molecules\cite{Singh2008}, or artificial atoms \cite{OConnell2010, Suh2010, Arcizet2011}, as they merge the robust and scalable infrastructure provided by NEMS/MEMS devices with the remarkable precision measurement and quantum control capabilities of atomic physics.

An impressive recent breakthrough in the study of quantum degenerate atomic gases is the ability to manipulate atoms trapped in optical lattices. This opens up a number of new possibilities in several frontier topics, including the control and quantum simulation of strongly correlated quantum systems \cite{Bakr2011, Simon2011, Endres2011, Cheneau2011} and quantum information. Alternatively, one could think of generating macroscopic cat states in mechanical systems \cite{OConnell2010}. As such it would be of considerable interest to transfer quantum states between ultra-cold atomic systems and mechanical oscillators, as this would offer an intriguing route to study the quantum dynamics of truly macroscopic systems and the quantum to classical transition. One particularly attractive aspect of quantum state transfer between micromechanical structures and atomic Schr{\"o}dinger fields is that both subsystems can have extremely low dissipation and decoherence rates compared to optical fields in resonators.

While there are now well understood optomechanical quantum state transfer protocols between optical and phonon fields and between electromagnetic fields of different frequencies \cite{Zhang2003, Tian2010, Khalili2010}, this is not yet the case for state transfer between Schr{\"o}dinger fields and phonon fields. This letter describes a scheme that achieves that goal for the case of single-mode fields in a hybrid system consisting of an atomic BEC trapped inside a Fabry-P{\'e}rot cavity with a suspended end-mirror or equivalent micromechanical analog. While most manybody states of interest in condensed matter physics involve multimode fields, achieving single-mode state transfer is an essential first step, and developments in cavity optomechanics are proceeding to the control of multimode fields in the near future. A key result of our analysis is the demonstration that under appropriate conditions our hybrid system can be described by an effective beamsplitter Hamiltonian with a quantum noise source due to the eliminated optical field. The beam splitter Hamiltonian is well known as a paradigm for state transfer between subsystems and its appearance for our hybrid system opens the door to state transfer between a BEC and a micromechanical element.

The interaction between the oscillating end-mirror, the (non-interacting) BEC and the single-mode intracavity field is described by the Hamiltonian
\begin{widetext}
\begin{equation}
H = \hbar \Delta_c \hat a^\dagger \hat a +i\hbar (\eta \hat a^\dagger - \eta^*\hat a)+\frac{\hat p^2}{2m_m} + \frac{1}{2}m_m\Omega_m^2 \hat q^2 - \hbar \xi \hat a^\dagger \hat a \hat q + \int dx \hat \psi^{\dagger}(x)\left [-\frac{\hbar^2}{2m_a}\frac{d^2}{d x^2} +\frac{\hbar g^2\hat a^\dagger \hat a}{\Delta_a} \cos^2(k x) \right ] \hat \psi(x) + H_{\rm d}.
 \label{Hamil}
\end{equation}
\end{widetext}
Here $\hat a$ ($\hat a^\dagger$) is the bosonic annihilation (creation) operator of the light field. $\hat p$ and $\hat q$ are the momentum and position operators of the mirror of mass $m_m$. $\eta=\sqrt{2P\kappa/\hbar\omega_l}\exp(i\phi_L)$ describes the external driving of the optical cavity, where $P$ and $\omega_l$ are the laser power and frequency respectively. $\omega_c$ is the cavity frequency, $L$ its length and $\kappa$ its decay rate.  $\xi=\omega_c/L$ is the optomechanical coupling constant. $\Delta_c=\omega_l-\omega_c$ is the detuning between the light field and the nominal cavity frequency (in the absence of optomechanical effect), and $\Delta_a=\omega_l-\omega_a$ is the detuning between the light field and the atomic transition, assumed large enough that the upper electronic state can be adiabatically eliminated. The second term in the square brackets describes the off-resonant dipole coupling between the condensate atoms and the intracavity light field, in the form of an optical potential of period $\lambda/2$, where $\lambda = 2\pi/k$ is the wavelength of the driving field. Finally  $\hat \psi(x)$ is the Schr{\"o}dinger field operator for the condensate of atoms of mass $m_a$, and $H_d$ describes the coupling of the optical field, the condensate and the optomechanical mirror to thermal reservoirs. In what follows we neglect the dissipation of the matter-wave and mechanical modes, as they are orders of magnitude slower than the optical decay rate.

The cavity field propagating along the x-axis can predominantly impart a photon recoil $2\hbar k$ to the initial zero-momentum cold atoms via Bragg scattering, and we assume that phase-matching limits the production of higher scattering orders. Restricting our analysis to one-dimension ($x$) for simplicity we may expand the Schr{\"o}dinger field as
\begin{equation}
\hat \psi(t) \approx \hat{c}_0 \psi_0(x) + \hat{c}_2 \psi_2(x),
\label{psi decomp}
\end{equation}
where $\psi_0=\sqrt{1/L}$ and $\psi_2=\sqrt{2/L} \cos{2kx}$, with $\hat{c}_0^\dagger\hat{c}_0+\hat{c}_2^\dagger\hat{c}_2 =N_a$, the total number of atoms. Substituting this form into the atomic part of the Hamiltonian (\ref{Hamil}) gives
\begin{widetext}
\begin{equation}
H_{\rm atom}=\frac{\hbar (2k)^2}{2m_a}\hat{c}_2^\dagger\hat{c}_2 +\frac{\hbar g^2}{2\Delta_a}\hat a^\dagger\hat a \left( \hat{c}_0^\dagger\hat{c}_0+\hat{c}_2^\dagger\hat{c}_2\right) +\frac{\hbar g^2}{\Delta_a\sqrt{8}}\hat a^\dagger\hat a\left( \hat{c}_0^\dagger\hat{c}_2+\hat{c}_2^\dagger\hat{c}_0\right).
\end{equation}
Assuming that depletion of the zero-momentum component of the condensate is small, we treat it classically via the replacement $\hat{c}_0, \hat{c}_0^\dagger \rightarrow \sqrt{N_a}$. Then neglecting unimportant constant terms, the total Hamiltonian becomes
\begin{equation}
H=\hbar \tilde{\Delta}_c \hat a^\dagger \hat a +i\hbar (\eta \hat a^\dagger - \eta^*\hat a) +\hbar \Omega_m \hat c_m^\dagger c_m + +\hbar \Omega_2 \hat c_2^\dagger c_2 +  \hbar \hat a^\dagger \hat a\left [-\xi_m(\hat c_m^\dagger + \hat c_m)+\xi_2(\hat c_2^\dagger + \hat c_2)\right ]
+ H_{\rm d},
\label{eq:Ham2}
\end{equation}
\end{widetext}
where $\tilde{\Delta}_c=\Delta_c+g^2N_a/2\Delta_a$ is the shifted cavity detuning due to the atomic medium, $\hat q = \sqrt{\hbar/2m \Omega_m} (\hat c_m + \hat c_m^\dagger)$, $\Omega_2 = 2\hbar k^2/m_a$ is the recoil frequency of the atoms, $\xi_m= \sqrt{\hbar/2m\Omega_m} \xi$, and $\xi_2=\hbar g^2\sqrt{2N}/(4\Delta_a)$.

The operator $(\hat c_2 + \hat c_2^\dagger)$ can be interpreted as the (dimensionless) ``position'' of the recoiled condensate side mode in Eq. (\ref{psi decomp}). Hence, the last non-dissipative term in Eq.~(\ref{eq:Ham2}) is an optomechanical term where the position of the recoiled condensate component is subjected to the radiation pressure of the intracavity light field. The Hamiltonian (\ref{eq:Ham2}) therefore describes the interaction of the light field with two oscillating mirrors, one real and one effective. The sign difference between the optomechanical coupling of the suspended mirror and the condensate results from the fact that while the mirror is pushed by radiation pressure, the atoms in the condensate can be either attracted to regions of high field intensity or of low field intensity, depending on the laser's detuning from the atomic transition, as apparent from the definition of $\xi_2.$

The Heisenberg-Langevin equations of motion are
\begin{eqnarray}
\label{eq:Oeqmotion}
\frac{d \hat a}{dt} &=&-\left (i\tilde{\Delta}_c +i \hat \Phi +\kappa/2  \right ) \hat a +\eta +\sqrt{\kappa} \hat a_{\rm in}, \\
\frac{d \hat c_m}{dt} &=&-i\Omega_m \hat{c}_m + i\xi_m \hat{a}^\dagger \hat{a},\\
\frac{d \hat c_2}{dt} &=&-i\Omega_2 \hat{c}_2 - i\xi_2 \hat{a}^\dagger \hat{a}.
\end{eqnarray}
where $\hat \Phi \equiv [-\xi_m(\hat c_m^\dagger + \hat c_m)+\xi_2(\hat c_2^\dagger + \hat c_2)]$
is the combined optomechanical phase shift of the recoiled condensate and the moving mirror.  If the optical field is treated classically, $\hat a  \rightarrow \langle  \hat a \rangle$, then the quantum states of the two ``mirrors'' are uncoupled, although their oscillation frequencies depend on a common classical intracavity intensity, which in turn depends on the expectation value $\langle \hat \Phi \rangle$ of the optomechanical phase shift. We stress that this implies that state transfer between the Schr{\"o}dinger field and the mirror, if possible at all, relies fundamentally on the quantum nature of the light field.

We now introduce the dimensionless position and momentum variables $\hat x_j=\hat c_j + \hat c_j^\dagger$ and $\hat p_j=i( \hat c_j^\dagger-\hat c_j )$, where $j=\{m,2\}$.
In order to  adiabatically eliminate the dynamics of the optical field we proceed by first linearizing the system of operator equations around the classical steady state, with
\begin{eqnarray}
\hat x_j &\rightarrow& \langle \hat x_j \rangle+\delta \hat x_j ,\nonumber \\
\hat p_j  &\rightarrow& \langle \hat p_j \rangle+\delta \hat p_j, \nonumber  \\
\hat{a} &\rightarrow& \langle \hat{a}\rangle+\delta \hat{a} ,
\label{linear}
\end{eqnarray}
and $\hat{a}^\dagger\hat{a}\approx\langle a^\dagger a\rangle+\langle \hat{a}^\dagger\rangle\delta \hat{a} +\langle \hat{a}\rangle \delta \hat{a}^\dagger$.
The equation of motion for the expectation value $\langle \hat a \rangle$ of the intracavity field is then
\begin{equation}
\frac{d \langle \hat a\rangle}{dt} =-i\Delta'\langle \hat a\rangle +\eta,
\end{equation}
with steady-state value $\langle \hat a\rangle_s=\eta/\Delta'$, where we have introduced the complex detuning $\Delta'=\tilde \Delta_c + \langle \Phi\rangle -i\kappa/2$,  which accounts for the optomechanical frequency shift. The fluctuations about the steady state are given in the usual input-output formalism by \cite{MilWalls}
\begin{equation}
\frac{d \delta\hat a}{dt} =-i\Delta' \delta \hat a -i\delta \hat \Phi \langle \hat{a} \rangle_s + \sqrt{\kappa} \hat a_{\rm in}.
\end{equation}
This equation can be formally integrated to give
\begin{eqnarray}\nonumber
\delta \hat a(t) &=& \delta \hat a(0) e^{-i\Delta't} - i\langle \hat a\rangle_s \int_0^t dt' \delta \hat \Phi(t') e^{-i\Delta'(t-t')} \\
&&+\sqrt{\kappa} \int_0^t dt'\hat a_{\rm in}(t') e^{-i\Delta'(t-t')} .
\label{eq:da}
\end{eqnarray}
For times long compared to $\kappa^{-1}$, and a cavity decay rate much faster that the inverse response time of both the effective and mechanical mirrors, the first term on the RHS of this equation decays to zero, and the operator $\delta \hat \Phi(t')$ can be evaluated at $t$. With this approximation,
\begin{equation}
\delta \hat a(t) = \delta \hat a(0) e^{-i\Delta't}+\frac{\eta(1-e^{-i\Delta't})}{\Delta'^2}\delta \hat \Phi(t) + \hat f(t),
\end{equation}
with $\hat f(t)$ being the last term in Eq.~(\ref{eq:da}). 
Since $\hat a_{\rm in}$ is a noise operator with  $[\hat a_{\rm in}^\dagger(t), \hat a_{\rm in}(t')]=\delta(t-t')$ (we take  the thermal photon number $n_{\rm th} = 0$ for optical frequencies) we have, for $t_1 < t_2$
\begin{equation}
[\hat f(t_1), \hat f^\dagger(t_2)]=e^{-i\Delta'(t_1 -t_2)}\left(e^{\kappa (t_1-t_2)/2}-e^{-\kappa t_2}\right),
\end{equation}
with a similar form for $t_1 > t_2$. This commutator vanishes rapidly over the characteristic time scale of the mirror dynamics for large $\kappa$, except for $t_1=t_2$. Over that longer time scale, $\hat f(t)$ can therefore be thought of as a delta-correlated noise operator as far as the mirror motion is concerned, with
\begin{equation}
[\hat f(t_1), \hat f^\dagger(t_2)] \approx \frac{\kappa}{[(\tilde{\Delta}_c+\langle \hat \Phi \rangle)^2+\kappa^2/4]}\delta(t_1-t_2).
\end{equation}
From now on we assume parameters such that the steady-state value of the phase shift is $\langle \Phi\rangle$=0. With this assumption and setting terms with $e^{-\kappa t/2}\rightarrow 0$, the linearization ansatz (\ref{linear}) results in the optomechanical interaction
\begin{eqnarray}
V_{\rm om} &\simeq& \frac{\hbar |\eta|^2\delta \hat\Phi(t)}{\tilde{\Delta}_c^2+\kappa^2/4}+  \left(\frac{\hbar \eta^*\hat f(t)}{-i\tilde{\Delta}_c+\kappa/2}+{\rm c.c.}\right)\delta \hat\Phi(t)\nonumber \\
&-&\frac{2 \hbar|\eta|^2 \tilde{\Delta}_c } {(\tilde{\Delta}_c^2+\kappa^2/4)^2}(\delta\hat \Phi(t))^2.
\end{eqnarray}
The second-order contribution can result in quantum state transfer between the suspended mirror and the condensate side mode. Specifically, consider the situation where the system is prepared in such a way that the shifted frequencies of the two oscillators $\Omega_m'$ and $\Omega_2'$ are equal, where
\begin{equation}
\Omega_j'=\Omega_j-\frac{4|\eta|^2\tilde{\Delta}_c\xi_j^2} {(\tilde{\Delta}_c^2+\kappa^2/4)^2},\quad j=\{m,2\}.
\end{equation}
In an interaction picture with the time variation due to the shifted frequencies of the two oscillator operators removed, under the rotating wave approximation, and neglecting constant terms, we then arrive at the total effective Hamiltonian describing the coupling between the mechanical oscillator and the BEC
\begin{widetext}
\begin{equation}\nonumber
H=-\hbar \frac{4|\eta|^2 \tilde{\Delta}_c\xi_2\xi_m}{(\tilde{\Delta}_c^2+\kappa^2/4)^2}(c_m^\dagger c_2+c_2^\dagger c_m)
+\hbar \left[\frac{\eta \hat f^\dagger(t)}{i\tilde{\Delta}_c+\kappa/2}+\frac{\eta^* \hat f(t)}{-i\tilde{\Delta}_c+\kappa/2}\right] \left [-\xi_m(c_m+c_m^\dagger)+\xi_2(c_2+c_2^\dagger)\right ] +H_d,
\end{equation}
\end{widetext}
resulting in the equations of motion
\[
 \frac{d}{dt}
\begin{bmatrix}
\delta x_2  \\
\delta p_2  \\
\delta x_m  \\
\delta p_m
\end{bmatrix}
=
\Omega_{\rm ST} \begin{bmatrix}
0 & 0 & 0 & -1 \\
0 & 0 & 1 & 0 \\
0 & - 1& 0 & 0 \\
1 & 0 & 0 & 0
 \end{bmatrix}
\begin{bmatrix}
\delta x_2  \\
\delta p_2  \\
\delta x_m  \\
\delta p_m
\end{bmatrix}
+\chi'
 \begin{bmatrix}
0  \\
-\xi_2  \\
0  \\
\xi_m
\end{bmatrix}
\]
where $\Omega_{\rm ST}=4|\eta|^2 \tilde{\Delta}_c\xi_2\xi_m/(\tilde{\Delta_c}^2+\kappa^2/4)^2$, and
\begin{equation}
\chi' = \frac{i2\tilde{\Delta}_c(\eta^*\hat f(t)-\eta\hat f^\dagger(t))+\kappa(\eta^*\hat f(t)+\eta\hat f^\dagger(t))}{\tilde{\Delta}_c^2+\kappa^2/4}.
\end{equation}
The above Hamiltonian has a beamsplitter form which produces periodic exchange of correlations between two subsystems, here the real and effective mirrors. The appearance of the beamsplitter Hamiltonian is the key result of this paper and follows from the quantum fluctuations of the cavity field. The coupling between the real and effective mirrors in our system is reminiscent of the Casimir force between two mirrors that arises from vacuum field fluctuations. In our case, however, there is no average net force between the mirrors but rather the cavity field fluctuations serve to dynamically exchange fluctuations in the quadratures of the two mirrors at the state transfer frequency $\Omega_{ST}$. The term proportional to $\chi'$ is a noise term due to random momentum kicks arising from cavity field fluctuations.  Since both the frequency of state transfer and the noise term depend on the same parameters, they must be chosen carefully to optimize state transfer.

As an illustrative example we consider an oscillating mirror of mass $m=6$~ng and frequency $\Omega_m=2\pi \times 16$~kHz forming the end mirror of a Fabry-P{\'e}rot cavity of length 195~$\mu$m and cavity decay rate $\kappa=2.6\times10^7~{\rm rad/sec}^{-1}$. Furthermore, the cavity is red-detuned from the driving field by $\Delta_c=0.1\kappa$, the pumping rate is $\eta =  3.9\kappa$. The cavity is filled with a small $^{87}{\rm Rb}$ BEC with $N_a=25,000$ atoms, and the incoming laser light is detuned $\Delta_a=-2\pi\times127$ GHz from the D1 transition line.

\begin{figure}[ht]
\begin{center}
$\begin{array}{cc}
\mbox{\bf (a)} &
\includegraphics[width=2.6in]{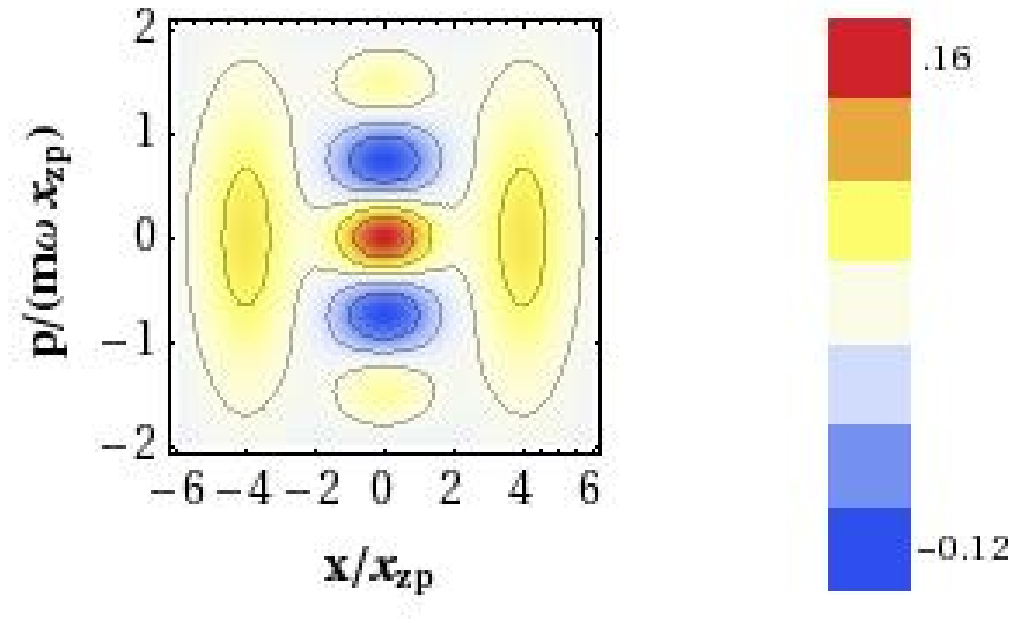} \\[0.0cm]
\mbox{\bf (b)} &
\includegraphics[width=2.6in]{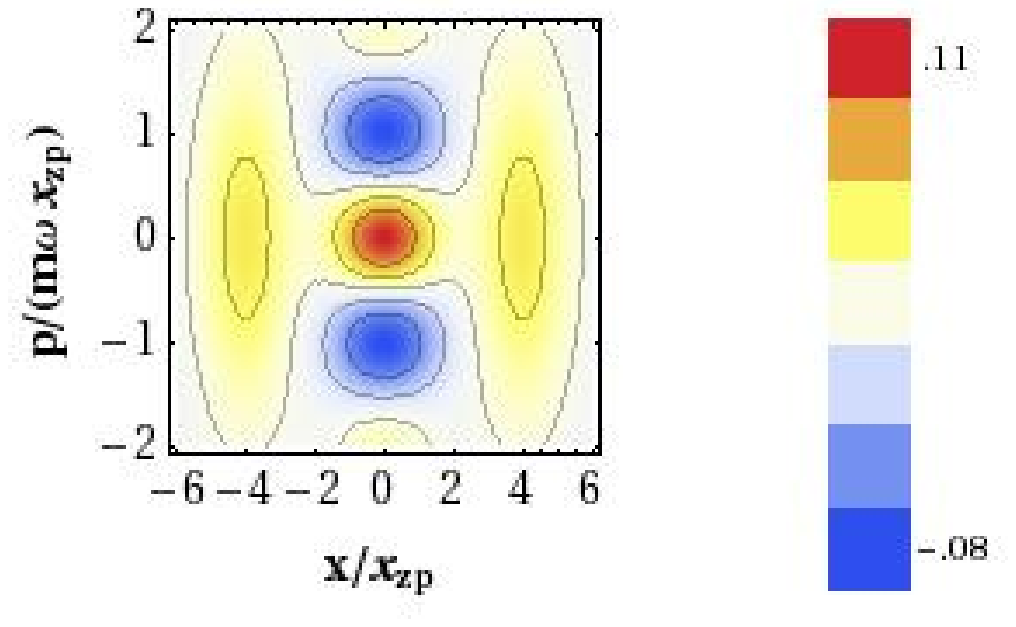} \\ [0.0cm]
\end{array}$
\end{center}
\caption{(Color online) Transfer of cat states: Wigner distribution functions of  (a) initial state of membrane $1/\sqrt{\mathcal{N}}(|\alpha\rangle+|-\alpha\rangle)$, where $\alpha=2$ in our dimensionless units, and (b) BEC after an interaction time of $t=\pi/(2\Omega_{\rm st})$. }
\label{fig:CatTransfer}
\end{figure}

Figure \ref{fig:CatTransfer} shows the results of our model for state transfer from the mechanical membrane to the BEC side mode for the case that the mirror is prepared in the Schr{\"o}dinger cat state $\frac{1}{\sqrt{\mathcal{N}}}(|\alpha\rangle+|-\alpha\rangle)$, with $\alpha=2/x_{\rm zp}$, $x_{\rm zp} =\sqrt{\hbar/m_m\Omega_m}$ being the width of the mirror ground state, and $\mathcal{N}$ being the normalization constant. The initial Wigner distribution $W(x_m,p_m,t=0)$ for the mirror is shown in the upper plot (a), and the corresponding Wigner distribution $W(x_2,p_2,t)$ for the BEC after state transfer at time $t=\pi/ 2\Omega_{\rm ST}$ is shown in the lower plot (b), and the similarity of the initial and transferred states indicates that the cat state nature of the initial state has been mostly preserved. In this example the quantum noise due to the adiabatically eliminated cavity field is not uniformly distributed between the two quadratures of the BEC field, and as a result the Wigner distribution of the transferred state in fig. \ref{fig:CatTransfer}(b) is stretched along the momentum axis in comparison to the initial state in (a) by a factor of $(\xi_2/\xi_m)^2$. As a consequence the fidelity of the state transfer (the magnitude of the overlap between the two Wigner functions), is 0.67.  Although this fidelity can be increased by parameter fine tuning or using a less complex initial state, the example serves to illustrate that state transfer is possible and also that the choice of parameters can significantly affect the fidelity of state transfer.


In summary, we have introduced a novel mechanism for state transfer between a trapped atomic gas and an optomechanical end-mirror that is mediated by the quantum fluctuations of a cavity field. We also presented the example of cat state transfer to illustrate the promise and pitfalls for high fidelity state transfer.  In future work we plan to extend of these ideas to multi-mode state transfer as appropriate to condensed matter systems, to add many-body effects, and to use quantum control and dark state approaches for improving the fidelity of state transfer.

We acknowledge stimulated discussions with L. Tian,  S.K. Steinke, L.F. Buchmann, A. Chiruvelli and H. Seok. This work is supported in part by the National Science Foundation,  ARO, and the DARPA ORCHID and QuASAR programs through grants from AFOSR and ARO.

\end{document}